# Anomalous interaction between dislocations and ultra-small voids

A. Dutta[1], M. Bhattacharya[2], P. Mukherjee[2], N. Gayathri[2], G. C. Das[1] and P. Barat[2]

*(1) Department of Metallurgical and Materials Engineering, Jadavpur University, Kolkata 700032, India*

*(2) Material Science Section Variable Energy Cyclotron Centre, 1/AF Bidhan Nagar, Kolkata 700064, India*

Amlan Dutta (Corresponding author)

Phone: +91 9903196791

Email: amlandutta2003@gmail.com

Formation of voids is a well known consequence of radiation damage in materials. These voids can pin down the motion of a moving dislocation and thus play a vital role in dictating the mechanical behaviour of crystalline soids. Initially, the strengthening effect of volume obstacles were studied using the line tension approximation for isotropic solids [1,2]. Later on, Scattergood and Bacon [3] developed an anisotrpic continuum model for dislocation void interaction taking into account the non-local dislocation self interaction, which was missing in the line tension approximation. Recently, the advent of atomistic simulation techniques has proved to be fruitful in studying the mechanism of dislocation pinning at voids at the atomic scales of length and time. In this context, the critical resolved shear stress (CRSS) is believed to be the most crucial parameter. This is defined as the minimum resolved shear load required to depin a given dislocation at a given void. The continuum model predicted that for a spherical void of diameter D and dislocation loop length L in a material with shear modulus G, the CRSS should be given by $\frac{Gb}{L}[\ln(D^{-1}+L^{-1})+\Delta]$, where b is the Burgers vector and $\Delta$ = 1.52. In their recent work, Osetsky and Bacon [4] reported that although the result of the contiunuum model fitted well to the outcome of the atomistic simulations for voids of moderate size, the CRSS were found to deviate from the expected values for voids of very small diameters. In the present work, we study the depinning of an edge dislocation in Tungsten at nanovoids of diameters on the order of ~1 nm using molecular statics simulations. As shown below, such small voids are found to exhibit features, which are quite unexpected from the conventional perspective of dislocation pinning.

Hatano and Matsui [5] reported the absence of thermal activation in the process of dislocation depinning at nanovoids in copper. Nevertheless, a reduction in the CRSS has been observed at elevated temperatures for both copper and iron during large scale molecular dynamics simulations [4]. Therefore it is desirable to perform the statics simulations, i.e. simulations at 0 K. Such simulations were performed in [4], where they revealed the pinning strength of individual voids, free from the effects of thermal assistance. Here we simulate the b.c.c. system using a many body interatomic interaction model [6], parametrized for Tungsten. An edge dislocation is introduced by splicing two crystals such that one of them is shorter than the other one by a single atomic plane and the system is left to relax. The dislocation line is along the $[1\bar{2}1]$ direction with its Burgers vector $0.3165[111]/2$ nm. Subsequently, spherical voids of diameters 0.8 nm, 1 nm, 1.2 nm and 1.4 nm were introduced at the centre of the simulation cell as shown in Fig. 1. The atoms in proximity of the defect site are identified by measuring their centrosymmetric deviations [7]. Periodic boundary condition (PBC) is imposed along all the three directions. As a result, two dislocations with opposite Burgers vector (not shown in the figure) are introduced above and below the primary dislocation. We now create an external shear stress in the simulation cell by implementing the Perinello-Rahman method [8] and relax the system to minimum energy configuration using the conjugate gradient relaxation algorithm [9]. We keep repeating the simulations with small increment in the shear stress at each run and relax the system for 2500 steps of iteration. At larger stresses, the dislocation line bows out and at large enough stresses, the dislcation is found to get depinned. Similar simulation strategy is often used to estimate the Peierls stress of dislocations. All the simulations are carried out using the MD++ molecular dynamics package [10].

It must be noted that the CRSS obtained in this method is somewhat overestimated and is not the true representative of the pinning strength of the obstacle, as the image effects are well known to contaminate the results in periodic simulation cells [11,12]. We can circumvent this problem simply by evaluating the critical stress for an obstacle free system under PBC and then subtarcting this stress (which turns out to be about 23 Mpa) from the depinning stress obtained in the simulations. This provides the *true* pinning strength by eliminating the effects of periodic images and the intrinsic Peierls stress. The typical response of the system to external shear stress is presented in Fig. 2. We immediately recognize that there are in fact multiple CRSS values for a given void size. This is clearly nontrivial as one expects depinning of the dislocation at all stresses exceeding the CRSS. In contrast, we observe that even if a dislocation is depinned at a given shear stress, it can remain pinned at a higher applied stress. As an example, we show the simulation snapshots for a void of diameter 1.4 nm as the system is relaxed under the shear stresses in Fig. 3. We find that the dislocation is depinned at 620 Mpa, whereas it is still pinned at 654 MPa.

In order to understand this surprising anomaly we must consider the underlying mechanism of dislocation pinning. In case of dislocation-void interaction, the inner spherical surface of the void provides free boundary condition and the terminating dislocation interacts with the image stresses. For plane free surfaces, the atomistic simulations [13] have already revealed that such image stresses can cause dislocation core relaxation at the terminating end, thereby exerting a pinning force. The continuum approach assumes the interaction of a dislocation with a spherical void. However, it can be noted that for a very small void, the structure is not spherical in true sense owing to the descrete crystalline structure of the solid. In addition, the applied shear stress has a tendency of distoring the shape of the void. Consequently, the atomic relaxation at the void surface alters with the stress and we obtain two processes acting simultaneously. Firstly, the applied load tries to depin the dislocation in the natural manner; secondly, it distorts the shape and thus modulates the pinning strength of the obstacle itself. These two effects are antithetical to each other and the dislocation gets depinned whenever the first effect is dominant over the second one, whereas it remains pinned if the second one predominates. This competition ultimately results in the occurance of multiple depinning stresses demonstrated in Fig. 2. To elucidate this point further, we create nanovoids in otherwise perfect crystals without dislocations. We add shear strains to the simulation cell in small incremental steps of $10^{-4}$ followed by relaxation of the system. The incremental shear strains are selected in a manner such that they create the shear stresses in a perfect crystal, which are very close to the applied shear stresses shown in Fig. 2. The total potential energy of the cell is recorded before and after implementing the relaxation algorithm and their difference provides the free energy of relaxation as shown in Fig. 4. As a matter of fact, the magnitude of relaxation energy is comparable to the strain energy increments due to the incremental shear strain of the solid and highlights the significance of this relaxation process. We also note that as a general trend, the amount of the released relaxation energy increases monotonically with the strain. This proves that the applied shear load tends to stabilize the structure of the nanovoid and hence raises the required depinning force. For the narrow ranges of shear strains studied here, the drop in relaxations energies appear to be linear and provide excellent linear fits as demonstared in Fig. 4. These fits produce good estimate of the strain sensitivities (change in relaxation energy per unit increment in shear strain), which are found to be -0.187 eV, -0.454 eV, -1.279 eV and -2.257 eV for the voids of diameters 0.8 nm, 1 nm, 1.2 nm and 1.4 nm respectively. On the whole, we note that these small voids can exhibit the tendency of releasing larger amount of the relaxation energy at larger strains (and stresses). Due to excessive relaxation of the atoms of the terminated dislocation core on the surface of a nanovoid, the dislocation effectively encounters an obstacle with enhanced CRSS at elevated shear stress. If the stress modulated CRSS exceeds the applied stress itself, the void can pin down the dislocation even if the dislocation is depinned at a smaller stress.

In conclusion, we find that the ultra-small voids interact with dislocations in quite unpredictable manner. As a matter of fact, there can be pinning gaps in between two depinning stress regimes and the applied load can itself enhance the strengthening effects of the nanovoids. The present work can motivate further studies of large scale systems containg a distribution of void dimensions. Similar studies can also be carried out to highlight the strengthening induced by ultra-small deposition of precipitates.

**Acknowledgement** One of the authors (AD) acknowledges the financial support provided by the CSIR, India.

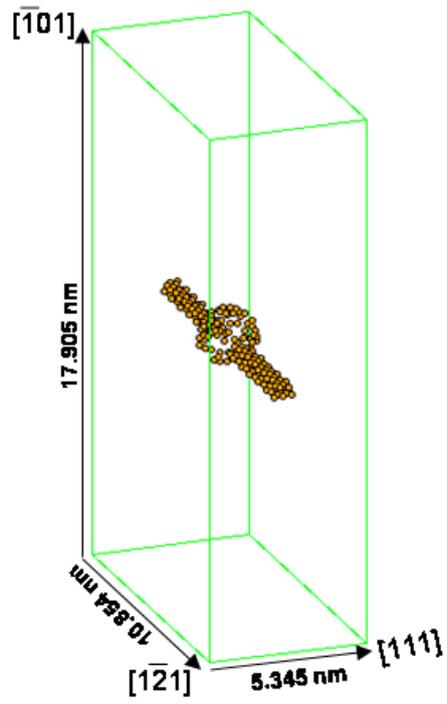

**Fig. 1** A typical simulation cell containing the edge dislocation and the nanovoid. Only the atoms belonging to the dislocation core and the nanovoid are displayed. The cell dimensions and crystal directions are also shown in the figure.

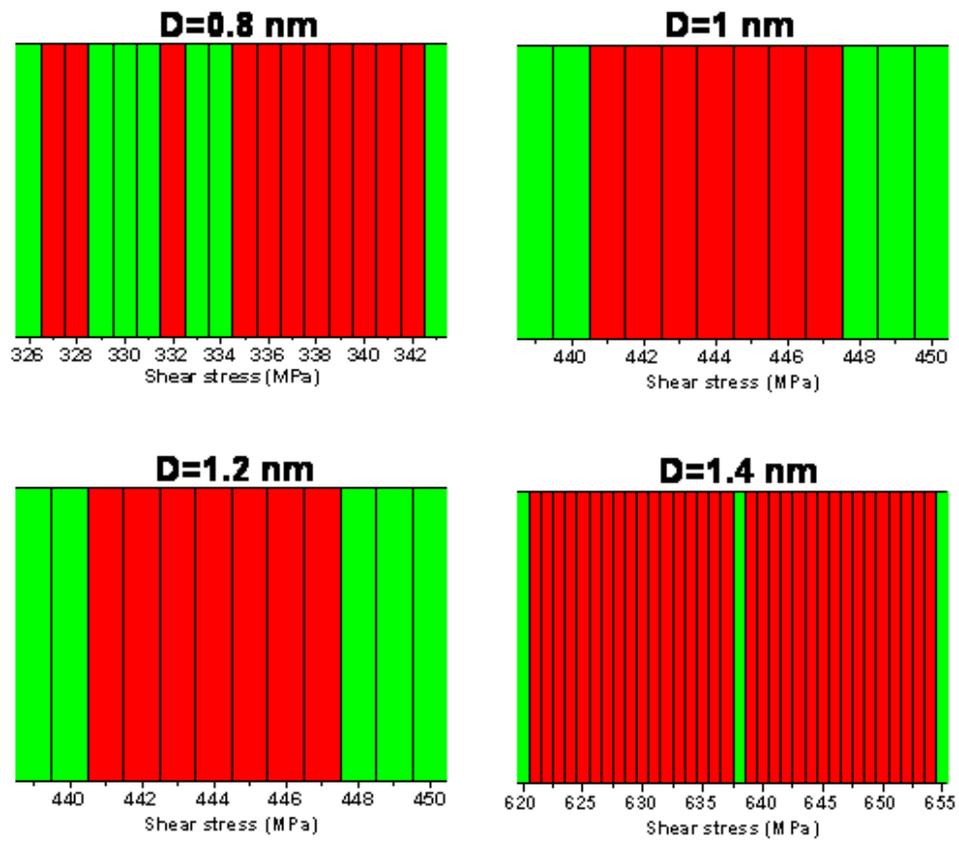

**Fig. 2** The response of the dislocation-void system to the incremental shear stresses for different void diameters *D*. The green bars indicate the stresses at which the dislocation gets depinned, while the stresses denoted by the red bars fail to depin it. One can observe that the *depinned regions* are seperated by the *pinned regions*.

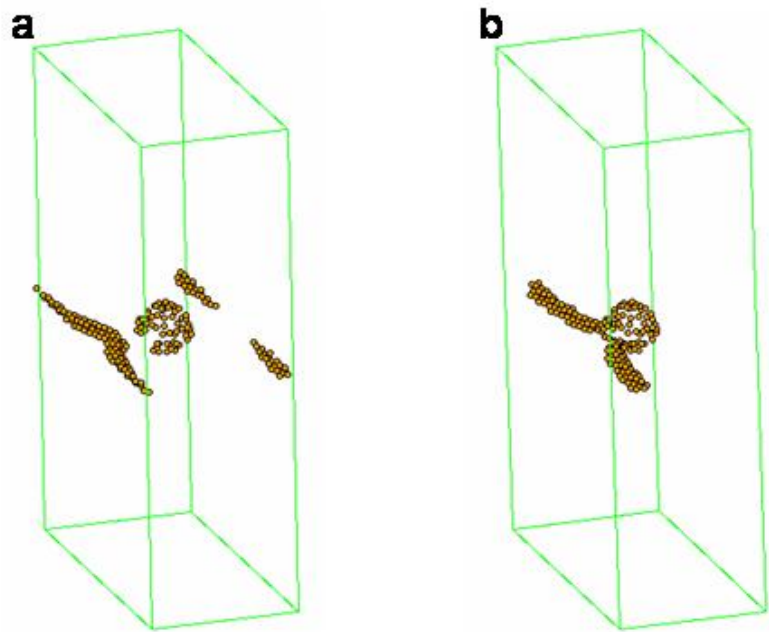

**Fig. 3** (a) Snapshot of the simulation cell when relaxed under a shear stress of 614 MPa. The dislocation is depinned and another dislocation form the neighbouring periodic image cell is entering the primary cell. (b) Simulation cell at 654 MPa. Despite a larger applied stress, the dislocation still remains pinned.

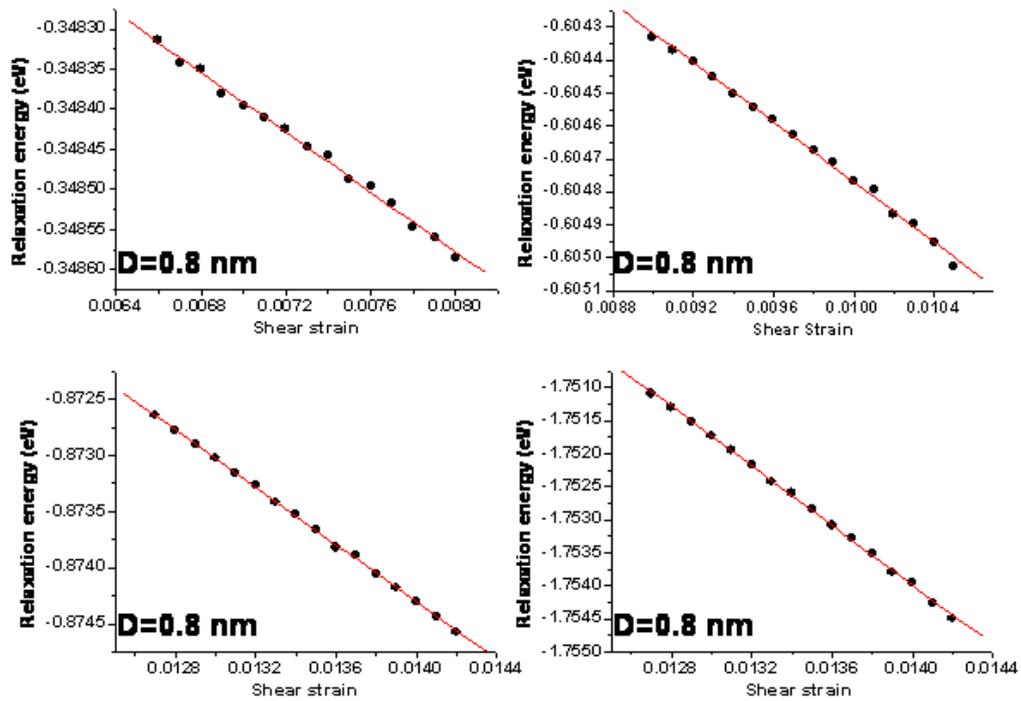

**Fig. 4** The free energies of relaxation for the nanovoids in a dislocation free crystal of dimensions 5.482 nm × 17.905 nm × 10.854 nm along the same crystal directions as shown in Fig. 1. The red line represents the linear fit to the scattered simulation data (filled circles).